\documentclass[conference]{IEEEtran}
\IEEEoverridecommandlockouts
\pdfoutput=1
\usepackage{cite}
\usepackage{amsmath,amssymb,amsfonts}
\usepackage{algorithmic}
\usepackage{graphicx}
\usepackage{textcomp}
\usepackage{xcolor}
\usepackage{hyperref}
\usepackage{array}
\usepackage[ruled,linesnumbered]{algorithm2e}
\usepackage{multirow}
\usepackage{enumitem}
\usepackage{makecell}
\usepackage{booktabs}
\usepackage{caption}
\usepackage{fancyhdr}

\def\BibTeX{{\rm B\kern-.05em{\sc i\kern-.025em b}\kern-.08em
    T\kern-.1667em\lower.7ex\hbox{E}\kern-.125emX}}

\begin{document}
\pagestyle{plain}

\newcommand{\eg}{\textit{e.g.},~}
\newcommand{\ie}{\textit{i.e.},~}
\newcommand{\aka}{\textit{a.k.a.}~}
\newcommand{\etc}{\textit{etc}}
\newcommand{\tl}[1]{{\color{red}[tl: #1]}}
\newcommand{\workname}{\textit{ClusterRCA}}


\title{ClusterRCA: An End-to-End Approach for Network Fault Localization and Classification for HPC System\\}

\author{
 \IEEEauthorblockN{
 Yongqian Sun\IEEEauthorrefmark{2}\IEEEauthorrefmark{5},
 Xijie Pan\IEEEauthorrefmark{2},
 Xiao Xiong\IEEEauthorrefmark{2},
 Lei Tao\IEEEauthorrefmark{2},
 Jiaju Wang\IEEEauthorrefmark{2},
 Shenglin Zhang\IEEEauthorrefmark{2}\IEEEauthorrefmark{6}\IEEEauthorrefmark{1}\thanks{* Shenglin Zhang is the corresponding author.},
 Yuan Yuan\IEEEauthorrefmark{3},\\
 Yuqi Li\IEEEauthorrefmark{3},
 Kunlin Jian\IEEEauthorrefmark{4}
 }

 \IEEEauthorblockA{
 \IEEEauthorrefmark{2}Nankai University,
 \{sunyongqian, zhangsl\}@nankai.edu.cn, \{panxijie, xiongxiao, leitao, wangjiaju\}@mail.nankai.edu.cn
 }

 \IEEEauthorblockA{
 \IEEEauthorrefmark{3}National University of Defense Technology, 
    yuanyuan@nudt.edu.cn, liyq@nscc-tj.cn
 }
 \IEEEauthorblockA{
 \IEEEauthorrefmark{4}Huawei, jiankunlin1@huawei.com
 }
 \IEEEauthorblockA{
 \IEEEauthorrefmark{5}Tianjin Key Laboratory of Software Experience and Human Computer Interaction
 }
 \IEEEauthorblockA{
 \IEEEauthorrefmark{6}Haihe Laboratory of Information Technology Application Innovation
 }
}

\maketitle

\begin{abstract}
Network failure diagnosis is challenging yet critical for high-performance computing (HPC) systems. Existing methods cannot be directly applied to HPC scenarios due to data heterogeneity and lack of accuracy. This paper proposes a novel framework, called \workname, to localize culprit nodes and determine failure types by leveraging multimodal data. \workname\ extracts features from topologically connected network interface controller (NIC) pairs to analyze the diverse, multimodal data in HPC systems. To accurately localize culprit nodes and determine failure types, \workname\ combines classifier-based and graph-based approaches. A failure graph is constructed based on the output of the state classifier, and then it performs a customized random walk on the graph to localize the root cause. Experiments on datasets collected by a top-tier global HPC device vendor show \workname\ achieves high accuracy in diagnosing network failure for HPC systems. \workname\ also maintains robust performance across different application scenarios.
\end{abstract}


\section{Introduction}
\label{sec:introduction}

High-performance computing (HPC) systems are essential for large-scale applications across various domains, such as weather forecasting, biological sciences, and fluid dynamics simulation \cite{usman2020big-hpc-importance}. These systems, typically organized into network clusters, facilitate complex computations and data processing. However, the intricate network structures of HPC systems make them susceptible to issues like network congestion and interconnect failures \cite{understanding}. Therefore, in the domain of HPC, accurate and efficient failure diagnosis is crucial for restoring network performance and ensuring uninterrupted operation. Failure diagnosis involves localization and classification\cite{FailureDiagnosisSurvey}, where localization helps narrow down troubleshooting by identifying the primary culprit node, and classification enables prompt determination of mitigation measures based on the specific failure type. Therefore, network failure diagnosis, including localization and classification, is critical for HPC system maintenance. 

When a task is observed to be faulty(\eg running timeout), the operator performs failure diagnosis on the components involved in this task, including root cause localization and type classification. Components in HPC systems typically include: compute nodes, network switches, storage systems, control nodes \etc. Among these components, the compute nodes and network switches are mainly responsible for performing the tasks, while network failures mostly occur among them. In contrast, failures on pure interconnected networks (\ie spine switches, without computing nodes) have been extensively studied and are relatively mature \cite{int} \cite{vargas2021peripheral}.
Therefore, in this work, we focus exclusively on network failures localization and classification at computing nodes and connected network switches(\ie leaf switches). In addition, we use ``node'' to refer to a compute node and network switch later in the paper.


\begin{figure}[ht]
	\centering
	\includegraphics[width=1.0\linewidth, height=0.15\textheight]{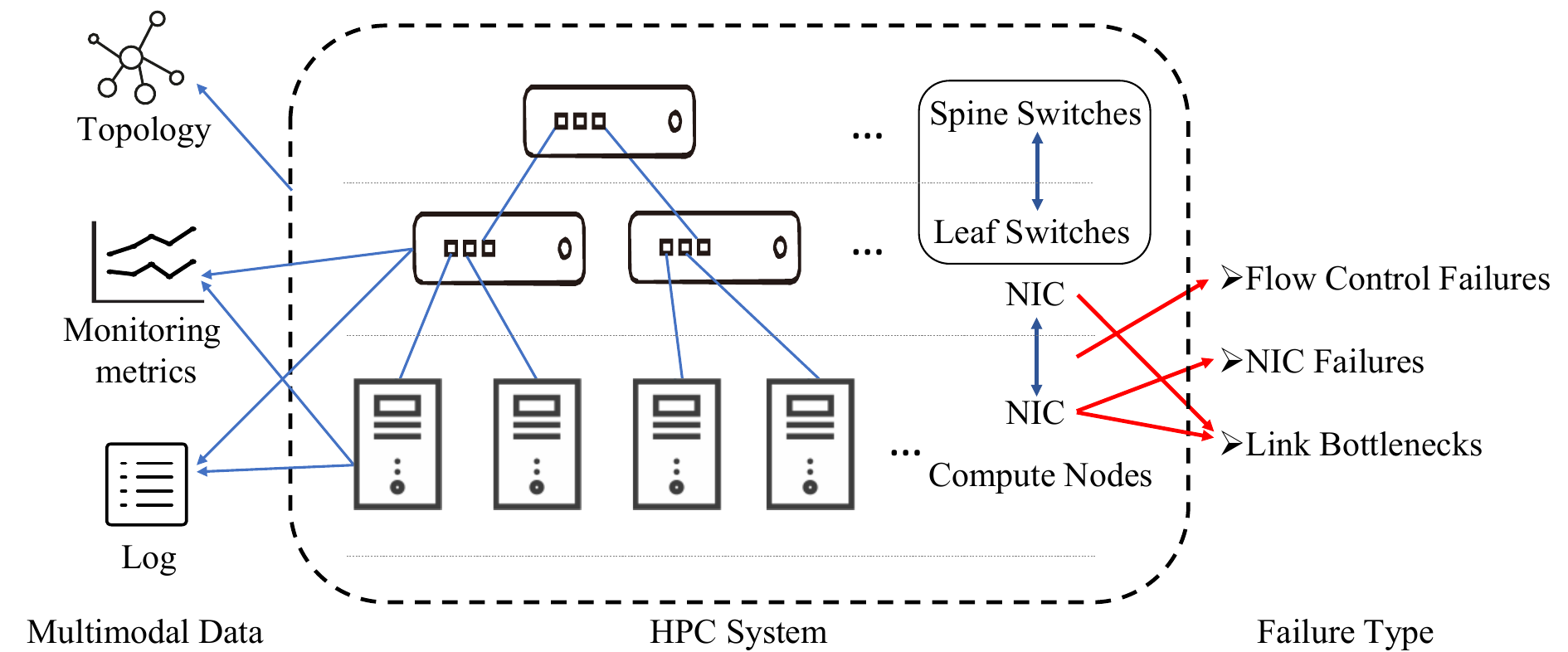}
	\caption{{Multimodal data and main types of network failure in an HPC system.}}
	\label{fig:HPCSystem}
\end{figure}

\begin{table*}[ht]
    \centering
    \caption{Detailed information of the network failures.}
    \label{tab:rootcause}
    \begin{tabular}{c|cccc}
        \toprule
        Category &Failure Type&Localization&Metric&Log\\
        \midrule
        \multirow{2}{*}{NIC failures}
        &Wrong packet&Computing node &rx\_crc\_errors\_phy $\uparrow$&CRC Error\\
        &PFC storm&Computing node &tx\_prio\_pause $\uparrow$&-\\
        \midrule
        \multirow{2}{*}{Link  bottlenecks}
	&Switch port performance restricted& Switch port &rx\_prio\_pause $\uparrow$&-\\
        &Tx timeout&Computing node &-&Tx Timeout\\
        \midrule
        \multirow{3}{*}{Flow control failures}
	&PFC parameter unaligned&Computing node  and Switch port &rx\_prio\_discards $\uparrow$&-\\
	&Switch PFC disabled&Switch port &rx\_discards\_phy $\uparrow$&-\\
	&Computing node PFC disabled  &Computing node &tx\_discards\_phy $\uparrow$&-\\
         \bottomrule
    \end{tabular}
\end{table*}

As shown in Fig. \ref{fig:HPCSystem}, the HPC system has the following characteristics:
\begin{enumerate}
    \item  The compute nodes and network switches are connected via a network interface controller (NIC) in a preconfigured topology. Synchronized communication between computing nodes is highly dependent on the interconnection network; network switches are indispensable to the network composition as necessary intermediate network devices connecting them. This topology correlates with the occurrence of failures in both.
    
    \item To monitor the network performance of HPC systems while ensuring that the systems deliver maximum arithmetic performance, the monitoring systems typically collect two types of monitoring data on the compute nodes and network switches during the execution of the application: logs and coarse-grained metrics (minute level). Logs are semi-structured text describing the devices' executing state, including state changes, software, and hardware errors, \etc. Metrics are multivariate time series, including CPU utilization, network I/O utilization, \etc. Besides metrics and logs, operators also document the topology of these devices.
    
    \item As listed in Table \ref{tab:rootcause}, different data modalities reflect various types of network failures. Analyzing data from just one modality is insufficient for precise failure diagnosis, as some faults are only reflected in metrics or logs.
     For example, due to the NIC sending queue of the computing node being full, network packet transmission timeout occurs, resulting in a large number of PFC messages. Or, the inability of the switching port to adjust the outgoing packet rate of the surrounding compute nodes results in the relevant compute nodes constantly receiving packets that exceed their processing capacity limits, at which point the metric ``rx\_discards\_phy'', which records the number of received packets that are discarded, will rise significantly.
    
    \item In our scenario, the monitoring data from compute nodes and network switches is heterogeneous. For example, we only need to record network resource metrics for the network switch, while for the compute node, we need to record additional CPU resource metrics. Similarly, it can be noted from Table \ref{tab:rootcause} that network failures will only occur on one of the components, except when the link connecting the two fails (\eg PFC parameter unaligned).
\end{enumerate}

In recent years, failure localization and classification methods can be categorized into two types: classifier-based~\cite{onlinediagnosis, cloud19, medicine, logcluster} and graph-based~\cite{sieve, causeinfer, automap, MicroCause}. Classifier-based methods attempt to directly distinguish the root causes from others by analyzing the features of monitoring data. However, the HPC system oriented approach~\cite{onlinediagnosis} is only for compute nodes and does not consider the heterogeneous data on the network switches and the correlation between the two. For other scenarios (\eg cloud and microservices)~\cite{cloud19, medicine, logcluster}, only a single modality of data is considered, which is limited in terms of the accuracy of the final diagnostic results.
Graph-based methods aim to establish causal relationships (\ie directed edges) between nodes to diagnose the root cause. However, most of these approaches focus on metrics, mining correlations between nodes through network traffic data or known service relationship information. In HPC scenarios, the collection of network metrics is often limited to a coarse-grained level to alleviate the computational resources occupied by the monitoring system, which, coupled with the lack of trace data that can directly represent dependencies, greatly increases the difficulty of graphical modeling.
Then later, some multimodal methods~\cite{cloudrca, ART, Eadro} combine the two. After using the classifier to unify the representation and feature extraction of multimodal data, the graph is then given to perform accurate root cause localization, which solves the problem of accuracy and the problem of difficult modeling.
Similarly, some methods~\cite{ART, Eadro} are applied in microservice scenarios, and the construction of the graph still requires trace or service relationship information. Another method~\cite{cloudrca} can only output the failure category and cannot be localized directly to a compute node or network switch.

Since both classifier-based and graph-based approaches have advantages in the context of HPC systems, we aim to explore combining these methods and their application in HPC systems through multimodal data to achieve more accurate failure diagnosis. However, in this process, we face two major challenges:


    
\begin{enumerate}
    \item \textbf{Device and data heterogeneity.} Because of how compute nodes and network switches are connected, it is impossible to separate them independently for failure diagnosis, which would ignore their correlation.
    
    
    
    \item \textbf{Combination of classifier and graph.} Data from two heterogeneous components makes classifier-based methods struggle to extract data features. Moreover, graph construction algorithms perform poorly on coarse-grained data (\eg PC algorithm\cite{MicroCause, automap}) or lack logical dependency data (\eg trace data\cite{surveyRCA2022, Eadro}, network traffic data\cite{sieve, causeinfer}). Previous methods lack accuracy in single diagnostic targets (\ie localization and classification) in HPC system scenarios, making it even more challenging to combine the two targets.
\end{enumerate}

To tackle the above challenges, we propose \workname, which integrates multimodal data for analysis, localizing the culprit node, and determining the failure type of network failures in HPC systems. For the special structure of the HPC system, \workname\ first translates logs and metrics into events through feature engineering, aggregating devices into NIC pairs based on the topological structure for unified modeling. To accurately localize culprit NIC pairs and determine failure types, \workname\ proposes integrating classifier-based and graph-based methods. First, it utilizes the output results of the classifier to establish directed graphs. Then, it adopts graph-based techniques for precise culprit NIC pairs localization and failure type classification by a customized random walk technique. Finally, based on the type of failure, it can determine whether the failure originated from a compute node or a network switch (or the link connecting them), so that the operator can carry out the corresponding remediation measures.

The primary contributions of our work can be summarized as follows:

\begin{itemize} 
\item To the best of our knowledge, \workname\  is a novel network failure diagnosis method capable of simultaneously localizing the culprit node and identifying the failure type using multimodal data in HPC systems.
\item We introduce a feature engineering technique based on NIC pairs. Utilizing the NIC pairs in the network topology, directly connecting the compute nodes and the network switches as the basic units, we unify the metric and log data features from heterogeneous equipment devices while taking into account the correlation between the two components, to address the challenge of conducting joint analysis posed by the complex structure of HPC systems.
\item To precisely localize the culprit node and determine the failure type, we propose a customized random walk technique with the help of a random forest. We use the classification results from random forest to guide the transitions in random walk, addressing the accuracy limitations inherent in standalone classifier-based and graph-based methods. {For greater reproducibility, our sample dataset and code are now publicly accessible.\footnote[
1]{https://anonymous.4open.science/r/ClusterRCA-F854}}
\item We conduct extensive experiments using data collected by a top-tier global HPC device vendor, evidencing \workname's diagnostic ability. {Specifically, results show that \workname\ achieves an average top-5 accuracy (Avg@5) of 0.9962, representing an average improvement of approximately 0.378 over the baseline methods, thereby demonstrating its superior network fault diagnosis performance.}
\end{itemize}

\section{Background}
\label{sec:background}
\subsection{HPC Systems and Monitor Data}
\label{sec:HPCSystem}
The network interconnects high-performance computing nodes and furnishes computing resources for HPC systems. Specifically, when a user executes a computing application on an HPC system, the application distributes the computing task to computing nodes (\ie clusters) and performs in parallel. During execution, computing nodes frequently communicate and synchronize through expeditious network messages with others to coordinate, facilitating the orderly processing of the computing task. This means network congestion or interrupts will affect the computing task or preclude continued execution. Consequently, HPC systems always equip network devices in clusters with a congestion control mechanism (Typically, Explicit Congestion Notification (ECN)\cite{ecn}) to sustain a fast and stable network. For clusters constructed on lossless networks, HPC systems always configure a flow control mechanism on devices (Typically, PFC in RDMA over Converged Ethernet v2 (RoCEv2)\cite{RoCEv2}) to avoid packet loss. The message traffic engendered by these mechanisms, together with the applied data traffic, is transmitted in vast quantities across the cluster network, increasing the complexity of the network environment. 

Computing nodes and switches constitute the primary nodes in the cluster, interconnecting via their NICs in a preconfigured topology. During application operation, they generate logs to record their network status, and their NICs record the type and count of each traffic traversing node. For each application, we collected three modal data (topology, metric, and log), distributed as shown in Fig.~\ref{fig:HPCSystem}. We describe their roles in network failure diagnosis below.

\textbf{Topology}: {In this paper, \textbf{topology} mainly refers to the connection structure between switches and computing nodes in the cluster.} When an HPC application executes, the computing nodes participating are selected, and the topology information indicates their interconnection edges within the cluster. Typically, a computing node first connects to a switch port and then to other computing nodes. Computing nodes have their NICs, Switch nodes have multiple NICs for each port, and NICs connect via a physical link. The interconnection between NICs constitutes topology, enabling operators to analyze the potential impact of relationships between nodes when network issues arise. Topology is recorded as a dictionary, \eg (``node'': ``server1'', ``nic'': ``NIC1'', ``link'': ``switch1'', ``link\_port'': ``100GE/18'') indicating that the NIC ``NIC1''  from the computing node ``server1'' interconnects with the 18th 100GE port of the switch ``switch1''. For each running application, there is a corresponding topology dictionary.

\textbf{Metric:} Monitoring metrics refer to the aggregated values of various packet types, including traffic packets \eg unicast and multicast packets, jumbo packets; loss or error packets \eg physical port discarded packets; traffic control packets \eg ECN packets received by the port; link quality packets \eg Cyclic Redundancy Check (CRC) error packets received by the port, and packets dropped due to physical encoding errors, \etc. These metrics are collected from the NIC registers on computing nodes and switches, separated into receive and send, and stored as time series. Furthermore, given that excessively frequent collection would occupy device computing resources, monitoring metrics are limited to the collection at large intervals (per minute in our scenario). When a network failure occurs, the operator calculates the difference between each count value and its previous value for every metric, examining changes in network traffic. Metrics intuitively reflect the congestion state of each node, enabling operators to deduce the rationale for their poor performance.

\textbf{Log:} Logs are semi-structured with timestamps and levels \cite{logsurvey}, recording software and hardware events from devices, such as software and hardware errors from computing nodes, the status of traffic control mechanisms, and configuration change events from switches, \etc. Operators frequently use the content and quantity information printed in logs to infer device operations and status. To efficiently analyze the vast number of logs generated by various devices, researchers have proposed several methods to automatically parse log information and extract it as templates and parameters\cite{drain, fttree, pop, spell}. 

\subsection{Categories of Network Failures} 
\label{sec:networkFailure}

 Network congestion and interruptions significantly impact HPC applications and remain an active area of research\cite{understanding, faultExascale, benchmark}. A stable HPC system network depends on the performance of cluster network devices, physical links, and flow control protocols. NICs are virtual network devices in HPC systems that analyze network traffic data. Proper NIC operation is crucial to cluster network performance. \textbf{NIC failures} arise from software and hardware failures, frequently causing incorrect parsing or message sending and inevitably affecting node communication with interconnected nodes. Therefore, NIC failure is one major category of root causes that lead to network failures, characterized by a large volume of error processing packets generated by the node's NIC, \eg improper packet errors and packet loss. Secondly, as physical links directly connect to nodes' NICs, uncontrollable environmental factors can cause \textbf{link bottlenecks}, preventing node traffic from meeting expectations and triggering many flow control mechanisms. \textbf{Flow control failures} refer to unsatisfactory flow control mechanism performance on cluster nodes. Flow control mechanisms are essential to maintain lossless network performance, and failure causes unprocessed flow control packets and packet loss due to excessive traffic. The types of root causes for network issues that we focus on are listed in Table~\ref{tab:rootcause}.

Even when originating at a single node in the cluster, the failures above will propagate expeditiously at a time far shorter than the traffic monitoring metrics collection interval. Failure symptoms (\ie packet errors, packet loss, and flow control messages) spread from the culprit node, obscuring the impact of the culprit nodes and victim nodes and generating complex failure patterns.  


\subsection{Problem Statement} 
\label{sec:problem statement} 
When a network failure occurs, operators need to localize the culprit node and determine the failure type to achieve timely failure mitigation. For HPC systems, the first task is a ranking problem: to rank the culprit node (computing node or switch port) higher than other nodes. The second task is a classification problem: classifying a network failure into a well-defined set of failure types. 


\section{Methodology}
\label{sec:algorithm}

\subsection{Design Overview}
\label{sec:overview}

\begin{figure}[ht]
	\centering
	\includegraphics[width=1.0\linewidth, height=0.21\textheight]{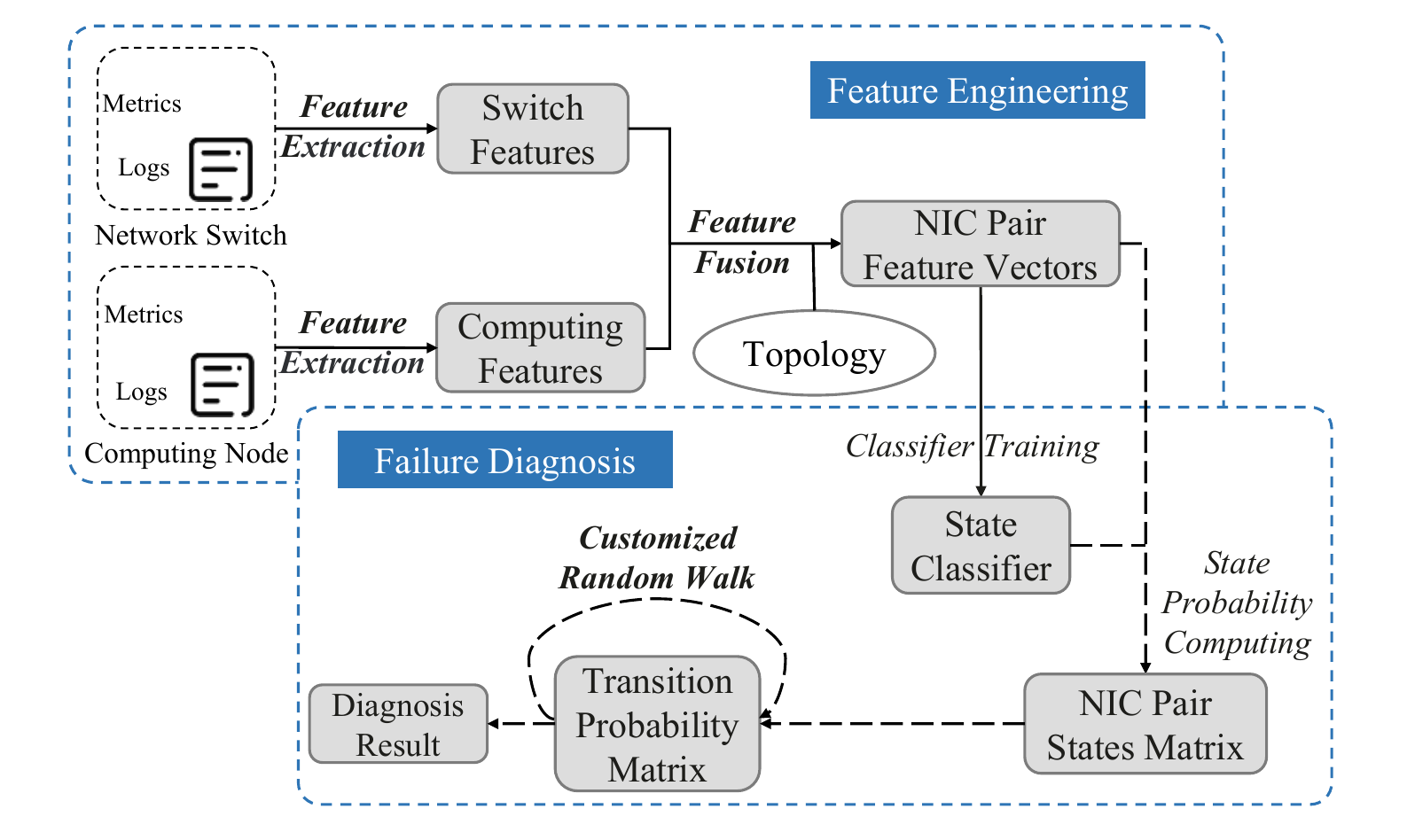}
	\caption{The overview of \workname. Solid lines denote processes that contain offline training and online diagnosis, and dashed lines denote only online diagnosis processes.}
	\label{fig:overview}
\end{figure}

This paper proposes a novel framework for localizing culprit nodes and determining failure types upon network failure in HPC systems. As shown in Fig.~\ref{fig:overview}, \workname\ comprises two phases: offline training and online diagnosis. 
The offline training phase will use historical metrics and logs to train feature extraction models and perform feature fusion based on historical topology to obtain the NIC pair feature vectors. This way, we can uniformly represent multimodal data while ensuring information integrity.
Then, the NIC pair feature vectors labeled as culprit, victim, or normal by the operator are input into the state classifier for training with corresponding labels. In this process, the state classifier learns various types of vector patterns to gain the ability to discriminate the data.
In the online diagnosis phase, real-time metrics, logs, and topology data are used to generate NIC pair feature vectors, which are fed into a trained classifier to obtain the state. Then, a state transfer matrix is constructed, which is combined with a state-specific transition strategy and a customized random walk algorithm to accurately localize the culprit node and determine the type of fault.

\subsection{Feature Engineering}
\label{sec:featureEngineering}

\begin{figure}[ht]
	\centering
	\includegraphics[width=0.8\linewidth, height=0.2\textheight]{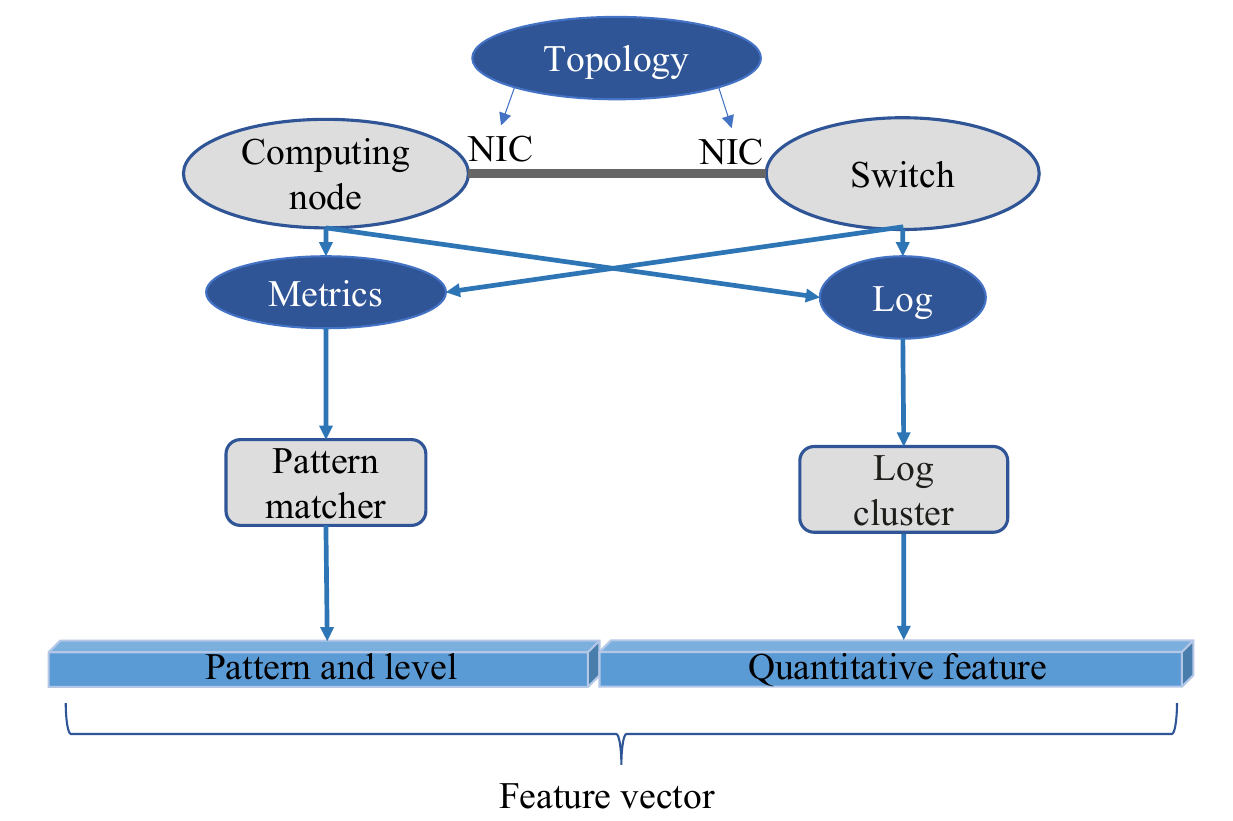}
	\caption{The feature extraction and fusion for a NIC pair.}
	\label{fig:NICpairs}
\end{figure}

In this subsection, we combine the NICs of the compute nodes with those of the switch ports directly connected to the compute nodes into NIC pairs. Then, we construct feature vectors for each NIC pair based on the network topology. Monitoring data for HPC applications is sliced into windows of defined time length (\eg 1 hour) and transformed into feature values. Fig.~\ref{fig:NICpairs} shows a typical situation: a NIC pair connecting a compute node and the specific switch port yields a feature vector after feature engineering. From the trained pattern matcher model and log cluster model, we extract pattern and level features from metrics and quantitative features from logs. All features are combined into a feature vector representing NIC pair characteristics for that window. 
After the feature engineering, the multimodal and scattered monitoring data in switches and compute nodes can be unified into a single NIC pair feature vector. We define the root cause candidate unit as a NIC pair feature vector with its failure type, thus solving the challenge of data diversity and device heterogeneity.

\subsubsection{Pattern matcher}
\label{sec:patternmatcher}

\begin{figure}[ht]
	\centering
	\includegraphics[width=1.0\linewidth, height=0.2\textheight]{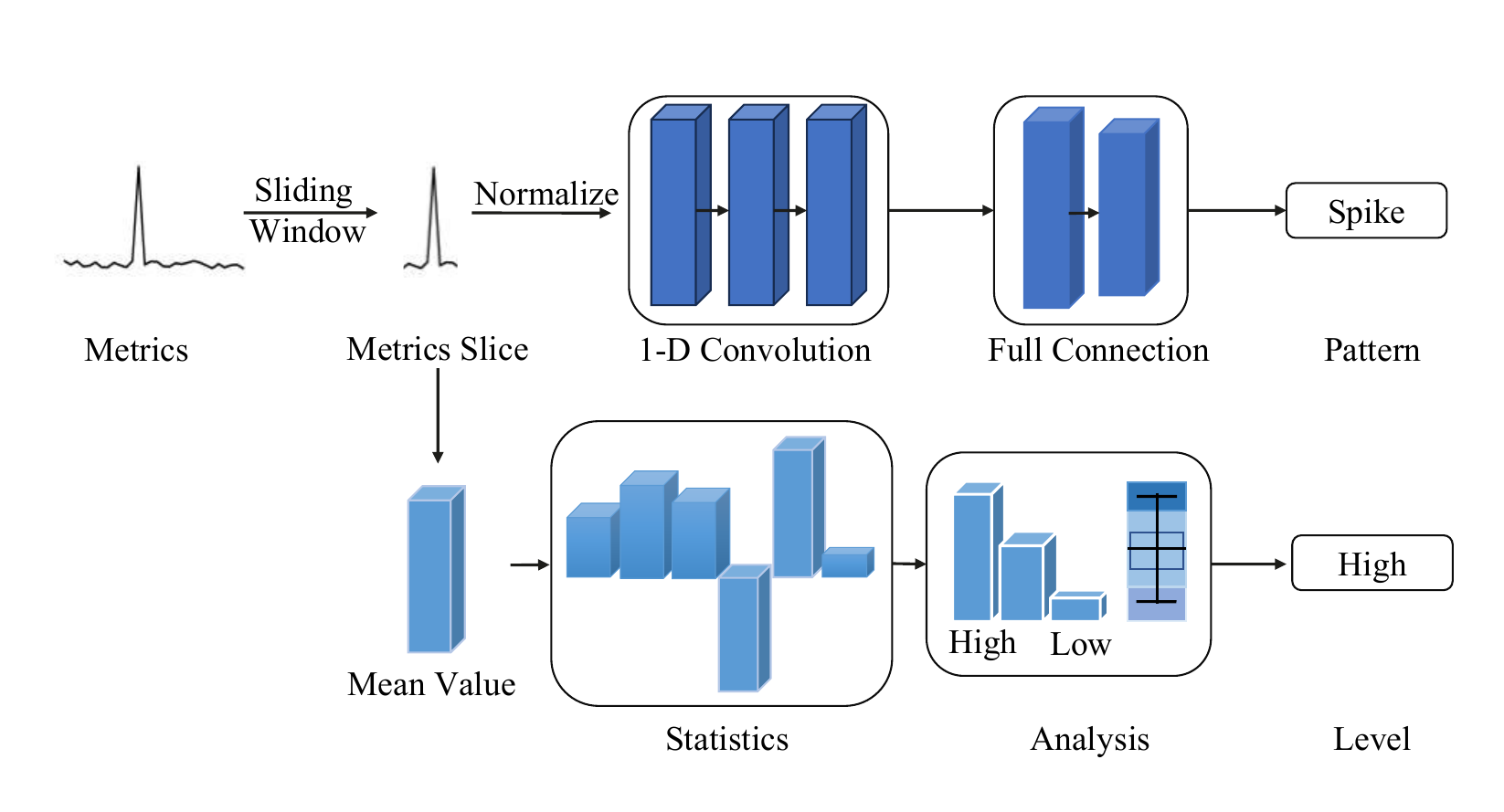}
	\caption{The feature extraction of metrics.}
	\label{fig:patternmatcher}
\end{figure}

Metric monitoring of HPC systems often has coarse time granularity to reduce resource consumption, which prevents millisecond-level traffic direction tracking \cite{patel2020uncovering}.
Therefore, valuable information primarily comprises metric change patterns and value levels. PatternMatcher~\cite{patternmatcher} mimics the visual perception of operators observing indicator curves and trains a one-dimensional neural network model to classify patterns of indicator changes that deserve attention (\eg steady rise, multiple peaks, sharp decline, \etc). Ultimately, each category will be mapped to a specific number. Fig.~\ref{fig:patternmatcher} (upper) shows the pattern matcher metric pattern extraction process. Inputs to 1D convolution and fully connected networks are normalized time-windowed metric slices. Outputs indicate pattern types. During offline training, operators label historical metric slices with patterns to train the neural networks. We asked an operator to label 1,000 metric slices in 6 hours. To reflect metric mean value levels, as shown in Fig.~\ref{fig:patternmatcher} (lower), we compute the mean of the same metrics across devices at the same time and use box plots \cite{boxplot} to determine levels and mark them with symbols.

\subsubsection{Log cluster}
\begin{figure}[ht]
	\centering
	\includegraphics[width=1.0\linewidth, height=0.12\textheight]{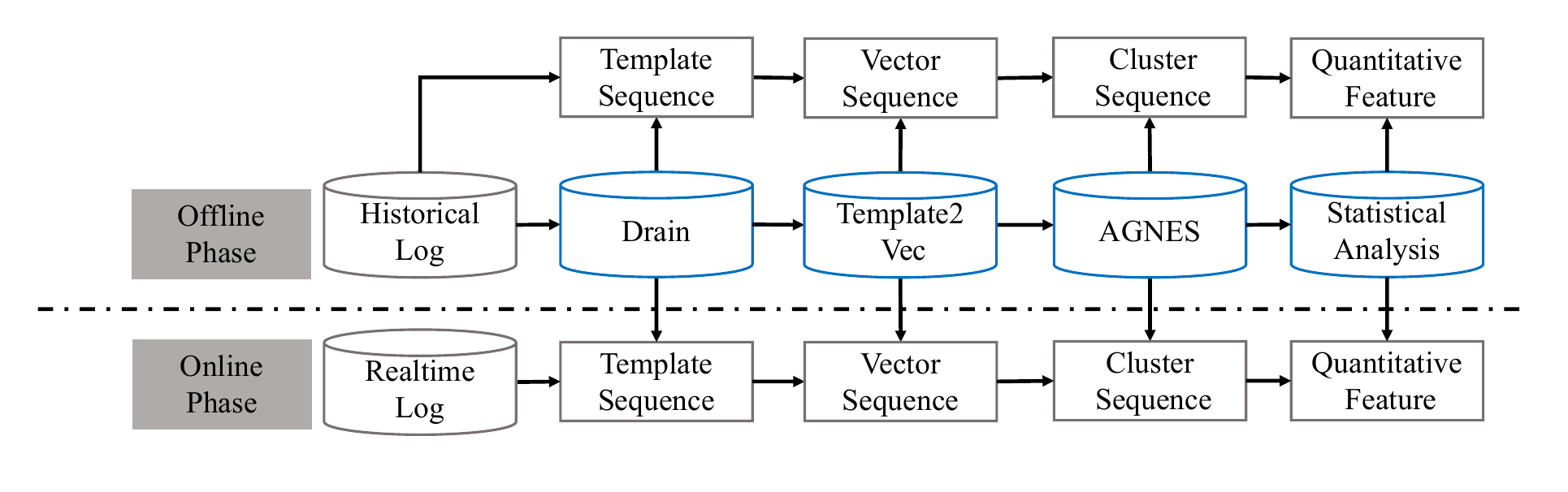}
	\caption{{The feature extraction of logs.}}
	\label{fig:logcluster}
\end{figure}

Some network failures also cause anomaly logs on the culprit and victim nodes. Anomaly log types and counts are critical features for distinguishing the root causes. 
{
The log feature extraction process is shown in Fig.~\ref{fig:logcluster}.
Firstly, we parse semi-structured logs into log templates using the Drain algorithm \cite{drain}. 
Next, these templates are converted into numerical vectors using Template2Vec \cite{template2vec} to effectively represent the template's semantic information numerically. 
To reduce dimensionality and group semantically similar events, we then apply the well-established AGNES (Agglomerative Nesting) algorithm \cite{AGNES} to these template vectors.
Finally, we analyze their numbers under normal conditions by computing cluster means and variances and using the 3-sigma principle to represent NIC pair log cluster quantitative features. 
In the offline phase, we retain the extracted log template, Template2Vec model, clustering details, cluster means, and cluster variances as model information for the online phase. }

Finally, we will concatenate the metric features and log features extracted from the same NIC pair. The feature data input into the subsequent process for fault diagnosis will include metric pattern features, numerical features, and log clustering quantitative features.

\subsection{Failure Diagnosis}
\label{sec:diagnosis}

In this subsection, we introduce the process of combining classifier-based and graph-based approaches, addressing the problem of failing to accurately identify the culprit nodes and classify the failure types using either the classifier or the graph approach alone. Specifically, we first use a state classifier (\ie random forest~\cite{randomforest}) to initially analyze the state probability of each NIC pair within the cluster belonging to the culprit, victim, and normal, then compute the transition probability between NIC pairs based on a specific transition strategy, and finally derive the diagnosis results through customized random wandering localization.

\subsubsection{State classifier}
Classifier-based diagnosis methods distinguish root causes from other states using well-trained classifiers, which is also helpful in our approach. We previously extracted node metrics and logs as features, but the combinations were too complex for effective classification. Therefore, we first apply anomaly detection for feature compression. 

{
Our observation that nodes with similar traffic patterns exhibit similar characteristics, with anomalies altering a few, is rooted in the inherent design of HPC systems. Identically configured nodes executing similar workloads naturally produce highly consistent operational behaviors, reflected in their performance metrics and log events. Anomalies, by definition, represent deviations from this expected consistency, thereby altering a subset of these features.

To capture features that are shifted due to anomalies, we compare features to normal nodes. We maintain a feature vector sample library that does not contain abnormal jobs and calculate the similarity between the real-time feature vectors and each normal sample in the library using Formula~\ref{formula:similarity}, where $V_r$ is the real-time feature vector, $M$ is the dimension and $V_{Normal}$ is the feature vectors in the normal sample library.

\begin{equation}
	\label{formula:similarity}
	similarity\left ( V_r, V_{Normal}\right ) =
	\frac{\sum_{M}^{i=1} 
	\mathbb{I} \left ( V_r\left [ i \right ]=V_{Normal}\left [ i \right ] \right ) }
	{M} 
\end{equation}
}

Since most features of the most similar normal sample match those of the sample, they can be considered to be in a very similar state. We compare the current sample with the normal sample position by position. Positions where there are differences between the sample and the normal sample are marked as abnormal values ``1'', and positions where there are the same are marked as normal values ``0''. In this way, the feature vector is converted into a 0-1 vector of the same dimension, simplifying the multi-dimensional feature space into a compressed representation of the anomalies that differentiate the sample from normal. In addition, if the current work is extended to another HPC network, it is only necessary to add feature vectors that represent typical data patterns to the normal sample library, without the need for retraining. Subsequently, this 0-1 vector is input into the state classifier.

\begin{figure}[ht]
	\centering
	\includegraphics[width=0.9\linewidth, height=0.17\textheight]{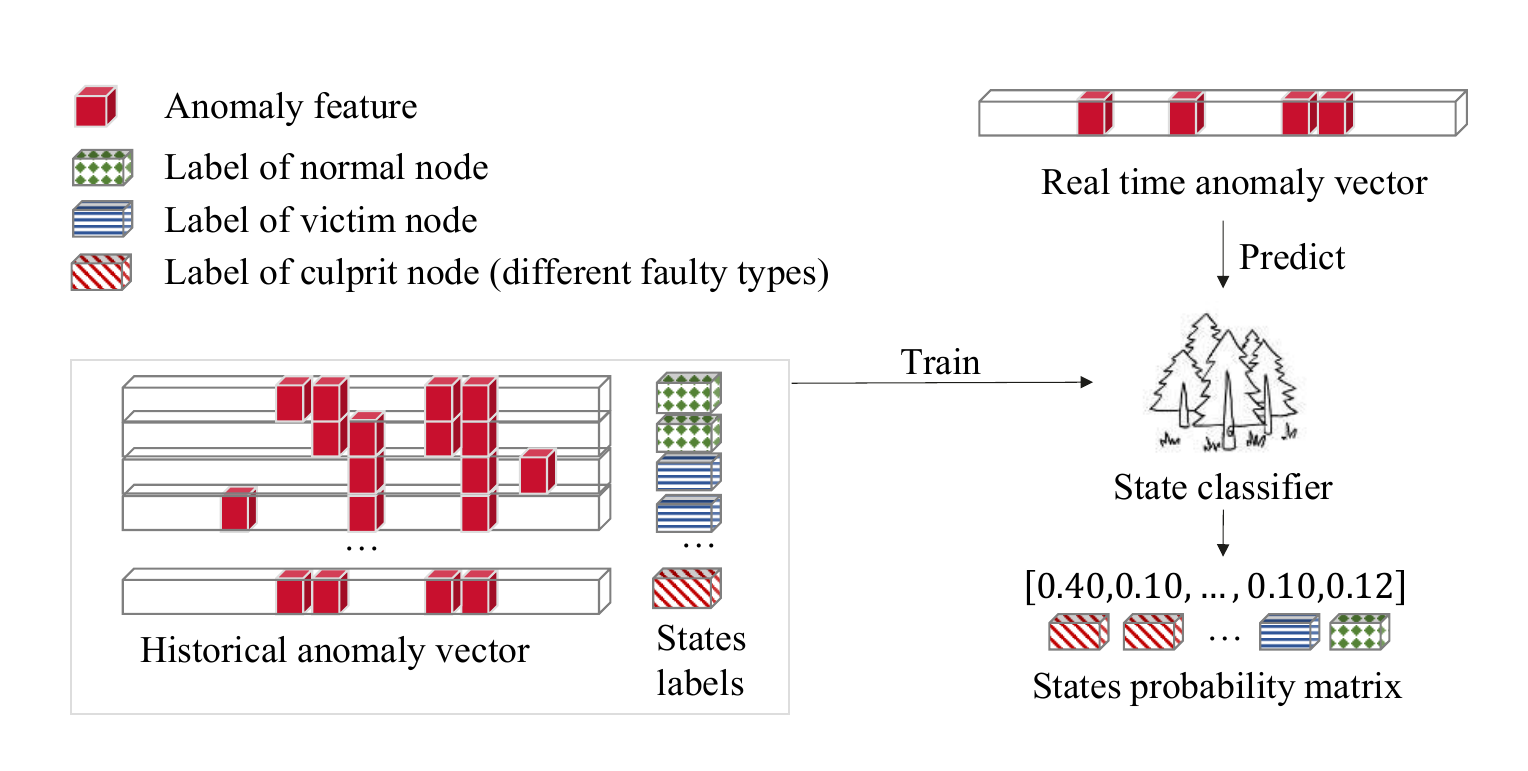}
	\caption{The training and predicting of the state classifier. }
	\label{fig:classifier}
\end{figure}

{
When performing fault injection, operators simulate a network failure on a specific node within a job, which is defined as the ``culprit'' node. Other nodes under this job are labeled as ``victim'' nodes, while the remaining nodes are classified as ``normal'' nodes. }

 As Fig.~\ref{fig:classifier} shows, we train a status classifier on historical NIC pair feature vectors and operator-provided labels  (nodes with various types of labels in Fig.~\ref{fig:classifier} are all NIC pairs). The well-trained classifier predicts a state probability matrix for a real-time feature vector, specifying the probabilities of nodes being the culprit of various failure types, a victim, or normal.

The multi-classifier used in this paper is the random forest \cite{randomforest}, which has been proven to outperform other classifiers in failure diagnosis for HPC systems \cite{onlinediagnosis}.
However, the results based on the classifier still need further analysis to determine the precise failure type conclusively or identify the culprit node uniquely, as there are typically multiple nodes exhibiting high probabilities of different failure types. Therefore, based on the preliminary results of the state classifier, we still need to use graph-based methods to determine the location of the root cause node.

\subsubsection{Customized Random Walk}
\label{sec:randomWalk}
Random walk is commonly used for failure diagnosis in cluster or cloud environments, as referenced in \cite{rootcauseincloud, randomwalk2018, automap}. However, these studies must first construct directional relationships between NIC pairs to develop transition strategies. Some studies on the application of the random walk method indicate that it can be used to improve the accuracy of the classifier by combining it with the classifier's results through special settings \cite{classificationbyRW, randomwalkreview}. Motivated by these studies, we propose a novel customized random walk with the help of the state classifier. Specifically, we utilize the classifier's probability matrix to formulate a state-specific transition strategy to initiate the random walk and then employ the random walk results to localize the culprit node that the classifier cannot conclusively determine. Specifically, the state classifier provides the probability matrix $S = [P_{N1}, P_{N2},...,P_{Nn}]$, where $P_{Ni}=[F_1, F_2,...,F_k, Victim, Normal]$, $F_k$ signifies the probability of NIC pair $N_i$ being the culprit node with $k^{th}$ failure type. $Victim$ and $Normal$ denote the probabilities of being the victim and normal nodes, respectively. 
Based on the above three types of probabilities, we define three states of the nodes and the corresponding transition tendencies as follows:

\begin{itemize}
\item \textbf{Culprit state:} Node $N_i$ is in the culprit state with probability $PF_{Ni} = P_{Ni}[F_1]+P_{Ni}[F_2]+...+P_{Ni}[F_k]$. If the walker is at the culprit node, it should remain, as this is the root cause. 
\item \textbf{Victim state:} When the walker reaches a victim node, it should proceed to the neighbor most likely to be the culprit. We set the probability of moving to neighbor $N_j$ to $P_{Ni}[Victim] * PF_{Nj} / \sum_{k\in B_{Ni}} PF_{Nk}$, where $B_{Ni}$ denotes $N_i$'s neighbors.
Note that since any two computing nodes of the HPC system can communicate, we fully connect all nodes (NIC pairs).
\item \textbf{Normal state:} Nodes in normal states show few anomalies, but the network may not be healthy. We want the walker to leave these nodes, randomly selecting neighbors quickly. The probability is $P_{Ni}[Normal] / |B_{Ni}|$, where $|B_{Ni}|$ signifies the number of $N_i$ neighbors. 
\end{itemize}

Based on the above three states, we define the transition probability matrix $Q$. Specifically, the probability of the current node $N_i$ transitioning to the adjacent node $N_j$ can be calculated as follows:

{\fontsize{6.5}{10}\selectfont
    \begin{equation}
        Q_{ij}=\left\{
        \begin{aligned}
        	PF_{i},\qquad \qquad \qquad\qquad &if\ j=i\\
        	\left (
        	\begin{aligned}
        		P_{Ni}[Victim] \times 
        		\frac{PF_{Nj}}{ {\textstyle \sum_{k\in B_{Ni}}PF_{Nk}} } \\ 
        		+ \frac{P_{Ni}[Normal]}{|B_{Ni}|}\qquad\quad
        	\end{aligned}\right ),& if\ j\neq i\ and\ N_j \in B_{Ni}\\
        \end{aligned}
        \right.
    \end{equation}
}

We then normalize each row of $Q_{ij}$ to obtain the final transition probability matrix. Each row of the matrix represents the probability of the walker transitioning from one node to another node in the graph, with the sum of these probabilities equaling 1. We can perform the rapid random walk with the transition probability matrix, as shown in Algorithm \ref{alg:randomwalk}.

\begin{algorithm}[ht]
    \small
	\SetAlgoLined
	\SetKwData{counts}{COUNTS}
	\SetKwFunction{RandomNode}{RandomInt}\SetKwFunction{RandomByPro}{ProbabilityBasedRandomChoose}
    \SetKwFunction{RandomUniform}{RandomUniform}\SetKwFunction{GetTrans}{GetTransitionProbabilityMatrix}
    \SetKwFunction{Max}{Max}
	\SetKwInOut{Input}{input}\SetKwInOut{Output}{output}
	
	\Input{number of results: $M$, number of walking steps: $STEPS$, list of states probability matrix: $S$, number of nodes: $N$}
    \Output{A ordered list of root cause: $result$}
	\BlankLine
		
	Initialize $result \leftarrow []$, start node $V \leftarrow$ \RandomNode{$N$}\;
	Initialize list of passed times \counts $\leftarrow [0] * N$\;
	Initialize transition probability matrix $Q \leftarrow$ \GetTrans{$S$}\;
	
	\For{$i \leftarrow 1$ \KwTo $M$}{
		\For{$j \leftarrow 1$ \KwTo $STEPS$}{
            $V \leftarrow$ \RandomByPro{$N$, $Q[V]$}\;
            \counts$[V]$ $\leftarrow$ \counts$[V]$ + $1$\;
		}
		$nodeId \leftarrow$ index of \Max{\counts}\;
		$faultId \leftarrow$ index of \Max{$S[nodeId][F_0:F_k]$}\;
		add $(nodeId, faultId)$ to $result$\;
		$S[nodeId][victim] \leftarrow S[nodeId][victim] + S[nodeId][faultId]$\;
		$S[nodeId][faultId] \leftarrow 0$\;
		
		$Q \leftarrow$ \GetTrans{$S$}\;
            \counts $\leftarrow [0] * N$;
	}
	\KwRet{$result$}\;
	\caption{Customized random walk with states probability}
	\label{alg:randomwalk}
\end{algorithm}

The algorithm initiates a random walk from an arbitrarily chosen initial node, traversing between nodes according to the transition probability matrix. It then determines the culprit node by checking the number of visits to each node. When the number of transitions exceeds the specified count $STEP$, the most visited node is the diagnostic culprit node. Its failure type with the highest state probability is the diagnostic failure type. The diagnostic result, which identifies the culprit node and failure type, is added to $result$. 

{Since the granularity of random walks is node-level, the final results must be refined to a more specific fault type level. If the root cause list is output based solely on the results of a single walk, it may overlook situations where multiple faults occur on a single node. Therefore, the algorithm converts the probability of the corresponding failure type for the last diagnostic culprit node to its victim state probability to obtain more combinations of culprit nodes and failure types. It then updates the transition probability matrix for the next iteration. After $M$ iterations, the algorithm returns an ordered list of $M$ root causes, each specifying the culprit node and associated failure type.}

\section{Evaluation}
\label{sec:evaluation}
Our evaluation answers the following research questions (RQs):\\
\textbf{RQ1}: How effective is \workname\ in network failure diagnosis?\\
\textbf{RQ2}: How robust is \workname\ in maintaining its diagnostic efficacy across diverse HPC application scenarios?\\
\textbf{RQ3}: Does each component of \workname\ have significant contributions to \workname's performance?\\
\textbf{RQ4}: How scalable is \workname in maintaining good performance on larger clusters?

\subsection{Experimental Setup}
\subsubsection{Dataset}

To evaluate the performance of \workname, we establish a small-scale HPC computing cluster environment in a top-tier global HPC device vendor ($\boldsymbol{A}$ company)'s laboratory.
The environment comprises four TaiShan server computing nodes and one network switch node. Each computing node connects to the switch node through a high-speed 100Gbps network. We deploy multiple HPC applications within this experimental setup and simulate a variety of network failures. 

As listed in Table \ref{tab:dataset}, we execute a total of six HPC applications spanning different domains:
\begin{enumerate}
    \item WRF\cite{wrf}: an HPC application developed by the National Center for Atmospheric Research in the United States for weather forecasting and climate research.
    \item Grapes\cite{Grapes}: a weather forecasting HPC application developed by the Chinese Academy of Meteorological Sciences since 2000.
    \item QE\cite{QE}: an HPC application developed by the Italian National Research Council for quantum chemistry calculations.
    \item GROMACS\cite{gromacs}: an HPC application for molecular simulations developed in collaboration with Uppsala University in Sweden.
    \item LAMMPS\cite{lammps}: an HPC application for molecular dynamics simulations developed by Sandia National Laboratories in the United States.
    \item OpenFOAM\cite{openfoam}: a C++ library for computational fluid dynamics developed by the University of Birmingham in the United Kingdom.
\end{enumerate}

Based on these HPC applications, we collect six datasets, including 739 network failure samples and 117 normal samples. We collect 176 network interface metrics and 38 computing resource metrics for each computing node and connected switch node port for each sample. The total number of log entries is 105,095. Each dataset contains samples of the failure types mentioned in Section \ref{sec:background}, namely \textbf{F1}: wrong packet, \textbf{F2}: PFC storm, \textbf{F3}: switch port performance restricted, \textbf{F4}: Tx timeout, \textbf{F5}: PFC parameter unaligned, \textbf{F6}: switch PFC disabled, and \textbf{F7}: computing node PFC disabled.

These network failures are simulated primarily through the injection of dedicated scripts. This approach allowed us to alter system-level parameters dynamically, reconfigure network device settings, and leverage specialized firmware utilities (e.g., Mellanox tools) to induce various fault mechanisms, such as error packets or periodic packet loss. Furthermore, we utilized packet generation tools and direct system calls to trigger specific queue behaviors, including Tx Timeout by stopping sub-queues. Throughout the execution of the HPC applications, we continuously collect monitoring data, including metrics, logs, and topology information from the computing and switch nodes.

The samples for various failure types in the datasets are listed in Table \ref{tab:dataset}. It is worth noting that due to the unique traffic characteristics of LAMMPS, it is almost unaffected by certain flow control error-related failures (no network congestion occurred). Consequently, we cannot collect valid samples for such failures in this application scenario. Furthermore, since the training process of \workname\ requires normal NIC-pair samples, we further divided the samples from the WRF\cite{wrf} application into WRF\_A and WRF\_B. WRF\_A contains all normal NIC-pair samples and approximately half of the failure NIC-pair samples used for training \workname, while WRF\_B is used to evaluate the failure diagnosis performance of \workname.

    
         

\begin{table}[ht]
    \centering
    \caption{Sample statistics in the data set.}
    \label{tab:dataset}
    \resizebox{\linewidth}{!}{ 
        \begin{tabular}{c|cccccccc|c}
            \toprule
             \multirow{2}[0]{*}{\centering Dataset}
             &  \multicolumn{2}{c}{\centering \thead{NIC\\ failures}}
             &  \multicolumn{2}{c}{\centering \thead{Link\\ bottlenecks}}
             &  \multicolumn{3}{c}{\centering \thead{Flow control\\ failures}} 
             &  \multirow{2}[0]{*}{\centering Normal}
             &  \multirow{2}[0]{*}{\centering Total}\\
             \cmidrule(lr){2-3}
             \cmidrule(lr){4-5}
             \cmidrule(lr){6-8}
        
             & F1 & F2 & F3 & F4 & F5 & F6 & F7&&\\
             \midrule
             
             WRF\cite{wrf} & 41 & 41 & 30 & 25 & 32 & 27 & 24 & 67 & 287 \\
             Grapes\cite{Grapes} & 24 & 19 &  24 & 19 & 23 & 24 & 24 & 10& 167 \\
             QE\cite{QE} & 31 & 18 & 22 & 11 & 29 & 16 & 25 &10& 162 \\
             GROMACS\cite{gromacs} & 14 & 13 & 13 & 12 & 13 & 14 & 10 &10& 99 \\
             LAMMPS\cite{lammps} & 10 & 10 & 10 & 11 & - & 7 & - &10& 58 \\ 
             OpenFOAM\cite{openfoam} & 13 & 14 & 14 & 14 & 9 & 4 & 5 &10& 83 \\
             \bottomrule
        \end{tabular}
    }
\end{table}

\subsubsection{Baseline methods}

We will discuss five representative baseline methods for comparison. Among them, AutoMAP \cite{automap} and Cloud19 \cite{cloud19} can only localize the culprit node; LogCluster \cite{logcluster} and CloudRCA \cite{cloudrca} can only determine the failure type; OnlineDiagnosis \cite{onlinediagnosis} can identify both the culprit node and the failure type. More details regarding these methods can be found in Section \ref{sec:related}.

Due to the different formats of these methods' analysis results, we only discuss the accuracy of the results they can output in the experiment. Specifically, we only access the accuracy of their culprit node localization for AutoMAP \cite{automap} and Cloud19 \cite{cloud19}. Conversely, the evaluation of LogCluster \cite{logcluster} and CloudRCA \cite{cloudrca} concentrates on determining the accuracy of failure type identification. For \workname\ and OnlineDiagnosis\cite{onlinediagnosis}, our evaluation encompasses accurately assessing identified culprit nodes and failure types.

\subsubsection{Evaluation metrics}

In the ordered failure diagnosis result list given by a diagnosis method, the higher the correct result ranks, the better the system effects. Therefore, we use \textit{top-k accuracy rate} ($AC@k$) and \textit{top-k average accuracy rate} ($Avg@k$) as evaluation metrics.

$AC@k$ calculates the proportion of the actual root cause test cases in the top $k$ of the ordered diagnosis result list. The higher $AC@k$ is, the more accurate the method is. Given a test set $A$, the calculation formula of $AC@k$ is shown in the Formula \ref{formula:ack}:
\begin{equation}
\label{formula:ack}
AC@k =
\frac{1}{|A|} \sum_{a \in A}\mathbb{I}({RC_i}_a \in {RC_s}_ak )
\end{equation}

${RC_i}_a$ is the actual root cause of test case $a$ (culprit node or failure type or both), and ${RC_s}_ak$ is the first $k$ results in the failure diagnosis result list of the diagnosis method.

$Avg@k$ is another evaluation metric that evaluates a method's overall capability of failure diagnosis, and the calculation is shown in Formula \ref{formula:avg}.
\begin{equation}
     \label{formula:avg}
     Avg@k=\frac{1}{k} \sum_{i=1}^{k} AC@ i
\end{equation}

Generally, operators only consult the first five failure diagnosis results, so we use $Avg@5$  to evaluate the average accuracy.

\subsubsection{Implementation}

All experiments are conducted on a Ubuntu machine server with Intel(R) Xeon(R) Gold 6138T Processor 2GHz CPU and 256GB memory.

\subsection{Overall Performance (RQ1)}
We first use WRF\_A as the training set and WRF\_B as the testing set to evaluate the overall performance of \workname\ and baselines. The two sets were collected when the same HPC application (WRF\cite{wrf}) was running. The accuracy of \workname\ and the baseline methods are listed in Table \ref{tab:overviewcpu}. Notably, the failure diagnosis capability of \workname\ significantly surpasses other methods, achieving a top-1 accuracy rate of 0.98. Cloud19\cite{cloud19}, limited by its search space, can only enumerate four root causes, resulting in a considerably lower $AC@1$ compared to \workname. LogCluster\cite{logcluster} solely considers log data; its $AC@1$ is limited to 0.20. In contrast, CloudRCA\cite{cloudrca} integrates information from logs and metrics data and achieves higher accuracy. OnlineDiagnosis\cite{onlinediagnosis} achieves considerable effectiveness ($Avg@5$ $>$ 0.95) because the statistical features extracted by OnlineDiagnosis accurately distinguish nodes' various network flow sizes. However, its accuracy will significantly decrease when the HPC application scenarios are inconsistent in the training and test sets. 

\begin{table}[ht]
    \centering
\caption{Performance of failure diagnosis.}
\label{tab:overviewcpu}
\resizebox{1.0\linewidth}{!}{ 
    \begin{tabular}{c|ccc|c|c}
        \toprule
         Method & $AC@1$ & $AC@3$ & $AC@5$ & $Avg@5$ & Time \\
        \midrule
         \textbf{\workname} & \textbf{0.9811}&\textbf{ 1.0} & \textbf{1.0} & \textbf{0.9962 } & 0.311s \\
         AutoMAP\cite{automap} & 0.2033 & 0.4390 & 0.6911  & 0.4488 & 0.355s \\
         Cloud19\cite{cloud19} & 0.5094 & 0.9245 & -  & - & 0.106s \\
         LogCluster\cite{logcluster} & 0.2075  & 0.4253 & 0.4253 & 0.3817 & 0.033s \\
         CloudRCA\cite{cloudrca} & 0.2830  & 0.7264 &\textbf{ 1.0}  & 0.6887 & 0.669s \\
         OnlineDiagnosis\cite{onlinediagnosis} & 0.8699 & 0.9756 & \textbf{1.0} & 0.9545 & 0.696s \\
         \bottomrule
    \end{tabular}
    }
\end{table}

\vspace{-5pt}

\subsection{Robustness Evaluation (RQ2)}

\begin{figure*}[]
	\centering
	\includegraphics[width=0.66\linewidth, height=0.48\textheight]{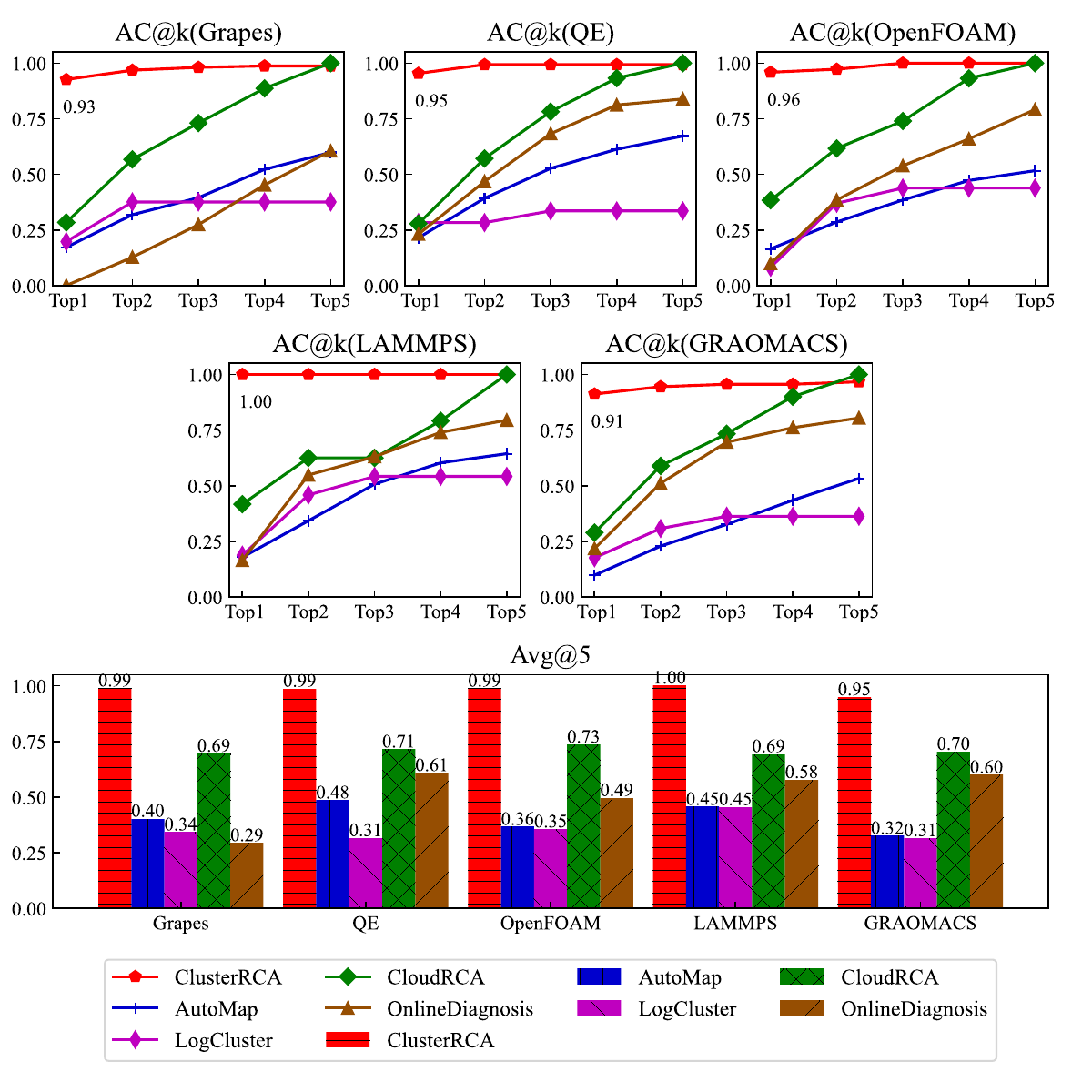}
	\caption{Result of robustness test.}
	\label{fig:generalization}
\end{figure*}

In practice, the applications running on HPC systems are diverse and constantly evolving, and the failure pattern caused by network failures also changes accordingly. For system administrators, running HPC applications is an uncontrollable user behavior. Therefore, we will analyze how \workname\ maintains performance across different HPC applications to determine whether it can adapt to the real network environment. Specifically, we use the WRF\_A  to train \workname\ and baseline methods. Then, we use test case sets of other applications except WRF\cite{wrf} to test these methods' failure diagnosis accuracy under different HPC applications. Cloud19\cite{cloud19}, offering only four root cause candidates, is excluded from this evaluation due to the unavailability of the $Avg@5$ calculation. The test case sets used for testing include Grapes\cite{Grapes}, QE\cite{QE}, GROMACS\cite{gromacs},  LAMMPS\cite{lammps}, and OpenFOAM\cite{openfoam}, and the results are shown in Fig. \ref{fig:generalization}. 

The first five subgraphs show that \workname\ can achieve $AC@1$ above 0.9 in five HPC applications, significantly higher than the baseline methods. The subgraph of $Avg@5$ shows that \workname\ has a significant failure diagnosis accuracy. Although OnlineDiagnosis\cite{onlinediagnosis} achieves an $Avg@5$ of 0.95 in the overall performance experiment, because the data feature extraction effect of OnlineDiagnosis\cite{onlinediagnosis} is more linked to the application data, under different HPC applications, its $Avg@5$ is only at most 0.73. Also, it can be observed that CloudRCA\cite{cloudrca} can reach the level of \workname\ in terms of Top5 accuracy because of more data features considered, but the results of the top root cause ranking are terrible. Both AutoMAP\cite{automap} and LogCluster\cite{logcluster}, on the other hand, are limited in their accuracy because that they only consider one data modality (metric or log data).

In summary, \workname\ can effectively determine the culprit node, clarify failure type across diverse HPC applications, and maintain a high accuracy to meet the needs of system administrators for network failure diagnosis. 

\subsection{Ablation Study (RQ3)}

In the ablation experiments, this work systematically removed some components of \workname\ and evaluated the performance of the remaining parts. By comparing the performance of the complete method and the ablation schemes, we can better understand the contribution of each component to \workname. First, we individually eliminated the features and data in the feature engineering part. Among them, we extracted two types of features for the metrics data: pattern features and level features. We deleted one of them respectively and consequently obtained two variants of \workname. Specifically, \textbf{C1:} Remove metrics pattern features; \textbf{C2:} Remove metrics mean level features. Then, we eliminated the input data.\textbf{ C3:} Remove metric data input;\textbf{ C4:} Remove log data input. Finally, we consider the random walk components. \textbf{C5:} Remove the random walk component. Deleting the entire state classifier component would cause the system to be unable to yield the failure diagnosis result list, which has not yet been tested.

\begin{figure*}[]
	\centering
        \includegraphics[width=0.67\linewidth, height=0.46\textheight]{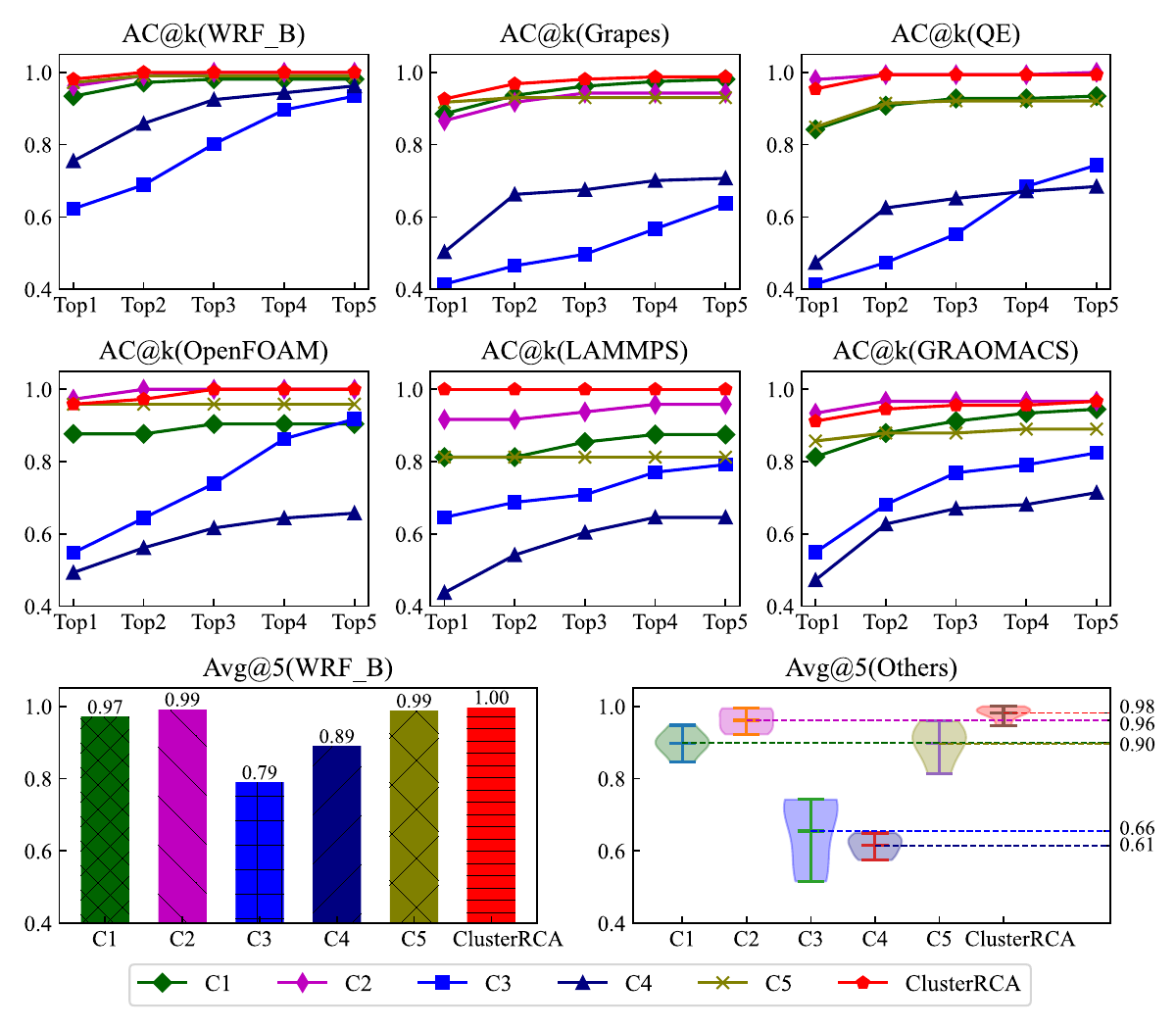}
	\caption{Contribution of components.}
	\label{fig:ablationcpu}
\end{figure*}

Fig. \ref{fig:ablationcpu} shows the experimental results of \workname\ and its variants, using the samples of WRF\_A for training and all our HPC applications for testing. \workname\ outperforms all variants in these tests. From subgraph $Avg@5$ (WRF\_B), it is apparent that the lack of input data (\textbf{C3} \& \textbf{C4}) significantly reduces the learning ability of the system, resulting in $Avg@5$ during training and testing on the same application (WRF\cite{wrf}) is much lower than that of  \workname\ and other variants. {Also, since different HPC applications have different traffic patterns and communication behaviors, the importance of metrics and logs varies. Therefore, in some applications, metrics alone perform better than logs alone, while in others, it is the reverse.} The subgraph of $Avg@5$ (Others)  shows the distribution of $Avg@5$ of \workname\ and variants across non-training application sample sets and marks the mean of $Avg@5$. Among them, the removal of some metrics feature types (\textbf{C1} \& \textbf{C2}) has a lower $Avg@5$ (0.90 \& 0.96) in other HPC applications than \workname\ (0.98), but not as apparent as directly removing the metrics data (\textbf{C3}, 0.66). It indicates that sufficient data types are more critical for \workname\ than abundant feature types. Comparing the average $Avg@5$ of \workname\ (0.98) and \textbf{C5} (0.90), it is evident that combining the random walk can achieve better diagnostic effectiveness than using the classifier alone. 

Overall, each component contributes positively to the final diagnostic performance, and together they enable \workname\ to achieve the best possible results. Metric and log data are the foundation of the model's ability to learn system behavioral patterns; removing either data causes the model to lose the ability to sense important information, especially metric data, which directly reflects the system's performance and health. 
Feature engineering helps the model capture failure-related patterns more effectively by distilling key information from the raw data; removing some feature types will affect the integrity of the information. 
The random walk component captures dynamic correlations or shifting patterns between different data sources, thus improving diagnostic accuracy.

\subsection{Evaluation on Scalability (RQ4)}

In real HPC systems, the dynamic allocation of computing resources by schedules determines the number and specific nodes for each job of HPC applications. Therefore, the failure diagnosis method must be feasible in practice, and the HPC cluster's dynamic topology must be considered, ensuring that the trained model exhibits robust scalability. Specifically, the failure diagnosis model trained on a small-scale cluster can also be directly used for a larger-scale cluster failure diagnosis while ensuring a certain accuracy. We collected two datasets with 16 and 32 nodes, respectively, in an HPC system simulation environment, and named them 16RANK and 32RANK (A is used for training, B is used for testing). As listed in Table \ref{tab:overviewgpu}, we test \workname\ on 16RANK\_B and 32RANK\_B, respectively, and use the small-scale data (16RANK\_A) to train \workname\ and evaluate its failure diagnosis ability on the larger-scale data (32RANK\_B). The results show that \workname\ has $Avg@5$ greater than 0.93 on both scales. The $Avg@5$ of \workname\ trained on 16RANK\_A is very close to that of \workname\ trained on 32RANK\_A (0.93 \& 0.94). The experiment proves that \workname\ also has excellent scalability.

\begin{table}[ht]
    \centering
    \caption{Scalability of \workname.}
    \begin{tabular}{cc|ccc|cc}
        \toprule
        
         Training&Testing& $AC@1$ & $AC@3$ & $AC@5$ & $Avg@5$ \\
        \midrule
         16RANK\_A&16RANK\_B & 0.9733 & 0.9867 & 0.9867 & 0.9840 \\
         32RANK\_A&32RANK\_B & 0.9419  & 0.9535  & 0.9535 & 0.9488 \\
         16RANK\_A&32RANK\_B & 0.9069  & 0.9419  & 0.9419 & 0.9349 \\
         \bottomrule
    \end{tabular}
    \label{tab:overviewgpu}
\end{table}


\section{Discussion}
\label{sec:discussion}

\subsection{{Observed Issues and Solutions}}
\textbf{Clock synchronization.} When a failure occurs, it propagates from the culprit node to other nodes, causing different failure patterns to appear on different types of nodes. If the clock difference between different nodes is too large, it may cause the key failure feature of this job to be lost. For example, the clock of node \textit{A} is relatively delayed by $\delta$, and the cluster starts to fail at $t_a$ (the time recorded by \textit{A} at $t_a-\delta$), while the time range for the diagnostic data is from $t_b$ to $t_c$. If $t_a-\delta$ is earlier than $t_b$, then the failure feature of node \textit{A} will be missed, and thus affect the subsequent construction of the failure propagation graph and inference. Therefore, the computing nodes and switch nodes in the cluster need to use clock synchronization algorithms to ensure that the monitoring data is aligned as much as possible. 

\textbf{Failure injection.} In our experiment, we inject failure by running scripts, and then construct corresponding failure labels (culprit node and failure type) for final performance evaluation. Then, for some unknown reason, sometimes failure injection may fail, without causing HPC performance degradation. In this case, the failure labels are incorrect. To ensure the accuracy of evaluation results, we will filter out the samples of incorrect failure injection according to specific rules and human verification.

\subsection{Threat to Validity}
\textbf{Data set limitations.} Compared to complex HPC systems, our cluster size is relatively small. Due to signing a confidentiality agreement with the $\boldsymbol{A}$ company, we cannot open-source all the original data of the experiment, as well as the fault injection scripts that involve system structure information. Although we also tried to find public network failure data sets related to HPC, we failed in the end due to the scarcity of HPC failure diagnosis research work. However, according to our experiment, \workname\ is effective and robust. It can also be applied effectively to larger HPC systems and more complex failure situations, which is very promising. 

\textbf{Data modalities limitations.} Our work involves using three multimodal monitoring data (metric, log, topology), while some HPC systems may lack metric or log collection. Since the feature extraction of metrics and logs is loosely coupled, \workname\ can still work normally in a data modality that is missing.

\section{Related Work}
\label{sec:related}
\textbf{Classifier-based diagnostic methods.} Some methods rely on accurate classifiers trained on historical failures to discern culprit nodes and failure types. For example, PatternMatcher\cite{patternmatcher} identifies metrics that trigger failures. It filters normal metrics, builds a pattern classifier to classify and filter unimportant metrics, and ranks the residuals by anomaly scores. OnlineDiagnosis\cite{onlinediagnosis} extracts statistics features from metrics, trains a random forest classifier to classify nodes in real-time failures, and ranks nodes by classification probability. Aside from metrics, Cloud19\cite{cloud19} and LogCluster\cite{logcluster} analyze log data. Cloud19 associates ERROR logs with nodes, uses Word2Vec to represent log entries, and trains a binary classifier to find the culprit node. LogCluster clusters logs by severity, trains a classifier to predict normal behavior, compares predictions and actual behavior to find the culprit node, and matches nodes to similar clusters. These methods use classifiers trained on metrics or logs, but single modal features cannot support accuracy, and excessive feature extraction (\eg OnlineDiagnosis's statistics features) leads to overfitting. 

\textbf{Graph-based diagnostic methods.} Graph-based diagnostic methods establish directed relationships between cluster nodes based on topology or system flow data. They then use random walks or causal inference to find culprit nodes. For example, MicroCause\cite{MicroCause} builds a directed acyclic graph showing metric dependencies using the PC algorithm and localizes the culprit node with an improved random walk. A similar method is AutoMap\cite{automap}. However, the PC algorithm must be improved to find cross-node metric propagation with multiple similar nodes. Sieve\cite{sieve} uses clustering to extract representative metrics for each node, combines business relationships into node dependency graphs, and makes causal inferences to determine the culprit node. However, HPC nodes lack business relationships. CauseInfer\cite{causeinfer} automatically builds node dependency graphs from network traffic data between monitored nodes and uses the PC algorithm for metric causal graphs within nodes. It then makes a two-layer causal inference to identify the culprit metric. Nonetheless, due to the high-performance computing demands, monitoring HPC network traffic at a fine granularity is not feasible.

\textbf{Combined Classifier and Graph diagnostic methods.} CloudRCA\cite{cloudrca} from Alibaba combines metrics and logs. It uses the PC algorithm to associate cluster features with failure types and a hierarchical Bayesian model to classify failure types, but not specific culprit nodes. ART~\cite{ART} designs a unified fault representation and automatically identifies anomalies based on extreme value theory thresholds, which is combined with cut-tree clustering to achieve unsupervised fault type triage, and cosine similarity to accomplish root cause localization.
Eadro~\cite{Eadro} jointly trains anomaly detection and root cause localization models through shared representations, and the loss function optimizes both tasks simultaneously, effectively using shared knowledge to reduce the negative impact of detection errors on localization.

In conclusion, classifier-based methods cannot achieve high accuracy for diverse HPC applications and monitoring data, and graph-based methods need help to construct accurate node relationships. Applying either alone cannot diagnose HPC network failures precisely.

\section{Conclusion}
\label{sec:conclusion}
Network failure diagnosis is of great importance for sustaining HPC system operability. In this paper, based on previous cluster failure diagnosis research, we first elucidate the challenges of culprit node localization and failure type identification from the heterogeneous data in HPC systems. We then propose a network failure diagnosis system, \workname. It leverages NIC-centered feature engineering to integrate metrics and logs from topologically adjacent HPC components, addressing the challenge of data diversity and device heterogeneity. Its failure diagnosis algorithm amalgamates classifier-based and graph-based methods so \workname\ can construct a directed graph to localize the culprit nodes and establish a highly accurate classifier to determine failure types. Finally, we evaluate \workname\ on experimental datasets from a top-tier global HPC device vendor ($\boldsymbol{A}$ company). All results demonstrate its robust diagnostic capabilities.

\section*{Acknowledgment}
This work is supported by the Advanced Research Project of China (No. 31511010501), the National Natural Science Foundation of China (62272249, 62302244), and the Fundamental Research Funds for the Central Universities (XXX-63253249).

\bibliographystyle{unsrt}
\bibliography{refer}



\end{document}